**Polariton-assisted donor–acceptor role reversal in resonant energy transfer between organic dyes strongly coupled to electromagnetic modes of a tuneable microcavity**

*Dmitriy Dovzhenko,\* Maksim Lednev, Konstantin Mochalov, Ivan Vaskan, Yury Rakovich, and Igor Nabiev\**

Dr. Dmitriy Dovzhenko, Mr. Maksim Lednev, Prof. Igor Nabiev
National Research Nuclear University MEPhI (Moscow Engineering Physics Institute)
115409 Moscow, Russian Federation
E-mails: dovzhenkods@gmail.com and igor.nabiev@univ-reims.fr

Dr. Dmitriy Dovzhenko
Department of Physics and Astronomy, University of Southampton, Southampton, SO17 1BJ, United Kingdom

Dr. Konstantin Mochalov, Mr. Ivan Vaskan
Shemyakin–Ovchinnikov Institute of Bioorganic Chemistry, Russian Academy of Sciences, 117997 Moscow, Russian Federation

Mr. Ivan Vaskan
Moscow Institute of Physics and Technology, Dolgoprudny, 141701 Moscow, Russian Federation

Prof. Yury Rakovich
IKERBASQUE, Basque Foundation for Science, 48009 Bilbao, Spain;
Donostia International Physics Center; Polímeros y Materiales Avanzados: Física, Química y Tecnología, UPV-EHU; Centro de Física de Materiales (MPC, CSIC-UPV/EHU) 20018 Donostia - San Sebastian, Spain

Prof. Igor Nabiev
Laboratoire de Recherche en Nanosciences, LRN-EA4682, Université de Reims Champagne-Ardenne, 51100 Reims, France

Resonant interaction between excitonic transitions of molecules and localized electromagnetic field allows the formation of hybrid light–matter polaritonic states. This hybridization of the light and the matter states has been shown to be able to significantly alter the intrinsic properties of molecular ensembles placed inside the optical cavity. Here, we have achieved strong coupling between the excitonic transition in typical oligonucleotide-based molecular beacons labelled with a pair of organic dye molecules, demonstrating an efficient donor-to-acceptor resonance energy transfer, and the tuneable open-access cavity mode. The photoluminescence of this hybrid system under non-resonant laser excitation and the dependence of the relative population of light–matter hybrid states on cavity detuning have been characterized. Furthermore, by analysing the dependence of the relaxation pathways between energy states in this system, we have demonstrated that predominant strong coupling of the cavity photon to the exciton transition in the donor dye molecule can lead to such a large an energy shift that the energy transfer from the acceptor exciton reservoir to the mainly donor lower polaritonic state can be achieved, thus yielding the chromophores' donor–acceptor role reversal or "carnival effect". Our experimental data confirm the theoretically predicted possibility for confined electromagnetic fields to control and mediate polariton-assisted remote energy transfer thus paving the way to new approaches to remote-controlled chemistry, energy harvesting, energy transfer and sensing.

**1. Introduction**

Strong light–matter coupling is a quantum electrodynamics phenomenon that takes place when the rate of resonant energy exchange (i.e., coupling strength) between the exciton transition in matter and the resonant localized electromagnetic field is higher than the competing decay and decoherence processes. Light–matter coupling then leads to the formation of two new "hybrid" light–matter (polaritonic) states with different energies, instead of the two original molecular and electromagnetic field energy states. Once the strong coupling regime is reached, the coupled system exhibits new properties possessed by neither the molecules nor the cavity [1]. Hence, by controlling the coupling strength, it is possible to modulate (or even control) various properties of the system, including the eigenenergy, excited state lifetime, efficiency and efficient distances of energy transfer, conductivity, etc. [2]. This paves the way to a wide variety of breakthrough practical applications, such as modification of chemical reactivity [3], enhanced conductivity [4], development of low-threshold sources of coherent emission [5], and even polariton simulators and logic [6,7].

Recently it has been demonstrated that strong coupling can modulate both distance and efficiency of Förster resonance energy transfer (FRET) [8,9]. FRET is a process of non-radiative energy transfer from one fluorophore (donor) to another one (acceptor). The FRET effect only occurs when several conditions are satisfied: (i) the donor emission spectrum should overlap with the acceptor absorption spectrum; (ii) the donor and acceptor fluorophores should be in a favourable mutual orientation, and, (iii) since the FRET efficiency is inversely proportional to the sixth power of the distance between the fluorophores, the distance between the donor and the acceptor should not exceed the Förster limit (10 nm).

When these conditions are satisfied, the FRET effect results in a decreased donor fluorescence emission accompanied by a simultaneously increased acceptor fluorescence emission.

Under the strong coupling regime, both donor and acceptor excitonic states could be coupled to the same microcavity optical mode, which may act as a mediator. This mediation can make it possible not only to increase the efficient distances of the energy transfer to values ten times larger than the Förster limit (to more than 100 nm) [9], but also to ensure an up to sevenfold increase in the rate of energy transfer [8]. This increase in the energy transfer rate leads to a significant decrease in donor fluorescence in the presence of the acceptor. As a result, the energy transfer efficiency, which is characterized by the ratio between the intensities of the donor fluorescence in the presence and absence of the acceptor, will be also significantly increased. It has been reported recently that under the strong coupling regime the energy transfer efficiency may be increased from 0.55 to 0.90 [8].

The possibility of increasing and controlling the FRET efficiency is promising for the development of many photonic applications, specifically for biomedical research. In this regard, one of the most powerful photonic nanotools are oligonucleotide-based molecular beacons, which are used in biosensing and specific gene identification [10-15], RNA imaging [14,16-18], revealing nucleic acid mutations [19], monitoring gene expression [20] and protein–protein interactions [21], nanomedicine [14,22], cell-surface glycosylation imaging [23], etc. All these applications are based on the same principle: a specifically designed molecular beacon is a circled oligonucleotide, with the donor and acceptor dye molecules conjugated to its ends and located in close vicinity to each other; hence, a strong donor-to-acceptor FRET occurs (**Figure S1** in the SI). Due to this strong FRET effect, the donor fluorescence is completely quenched and, in this respect, the beacon is invisible. When the beacon oligonucleotide binds to its complementary oligonucleotide target, its circled structure is opening, the distance between the donor and acceptor increases, and this binding event is reported by the appearance of the originally quenched donor fluorescence signal. The possibility of controlling the FRET efficiency within the molecular beacon by making the originally invisible molecular beacons visible "on demand" may extend their applications. One of promising ways to control the FRET efficiency is to use light–matter interaction in various regimes [8,24].

Polariton-assisted energy transfer between spatially separated molecules has been extensively studied in various configurations [8,9,25]. In Ref. [25], hybrid polaritonic states have been demonstrated to be an efficient energy transfer pathway between two spatially separated J-aggregates with an initially negligible direct energy transfer via dipole–dipole coupling. However, for many practical applications that we have mentioned above, it is even more intriguing to have a way to alternate the energy relaxation in a mixed media where the donor and acceptor molecules are located in close vicinity to each other and the FRET effect is mediated by direct dipole–dipole coupling between them. In Ref. [26], hybridization of the light and the matter states in a microcavity filled with a blend of two BODIPY fluorescent dyes with similar properties has been investigated. Here, the electromagnetic microcavity modes of light have been found to be coupled, at the same extent, with both donor and acceptor exciton transitions. It has been shown that, in this system, direct dipole–dipole coupling is more efficient than the energy transfer via strong coupling. Interestingly, the strong coupling to only one of the two excitonic states of the system has also been shown to be promising if the control over relaxation pathways in FRET systems has to be obtained. In Ref. [27], the authors have theoretically demonstrated that, whereas exclusive strong coupling of the cavity photon to the donor states can enhance the energy transfer to the acceptors, the reverse is not true. On the other hand, it has been shown that sufficiently strong coupling exclusively to the acceptor can modify the energy levels of the system in such a way that the transfer from the acceptor to the donor states mediated by polariton states can occur, leading to the chromophore role reversal or "carnival effect" [27]. It is interesting to investigate the possibility of combining these effects by means of strong coupling of exclusively the donor state to the cavity photon with a large coupling strength. This should lead to the formation of a polariton state with a relatively large fraction of the donor exciton and a small fraction of the acceptor exciton, with the energy lower than that of the original acceptor state. However, such a modification will require very strong coupling to ensure significant alteration of energy levels.

To date, most FRET studies using the strong coupling regime have employed only simple Fabry–Perot microcavities, which have relatively large mode volumes and, hence, rather moderate light–matter coupling strengths, which have significantly limited the experimental observations of the theoretically described effects [27]. Recently, we have engineered a tuneable microcavity with a lateral mode localization characterized by a drastically decreased mode volume, which allows obtaining a considerably larger coupling strength [28], and have employed this microcavity in the present study. In order to ensure a sufficiently strong coupling of the microcavity photon modes exclusively to the molecular beacon's donor state, we used the 6-carboxyfluorescein (FAM) dye, which has a large dipole moment and a high quantum yield as a donor, and the rhodamine derivative carboxytetramethylrhodamine (TAMRA) with much lower dipole moment and quantum yield as a molecular beacon's acceptor (see the *Sample preparation* section and the Supporting Information (SI) for details). The distance between these donor and acceptor dye molecules conjugated with the opposite termini of the molecular beacon was determined by the diameter of the DNA double helix (about 2 nm), which was short enough to ensure efficient direct dipole–dipole coupling between them (see **Figure S1** in the SI). This system made it possible to investigate the polariton-assisted energy transfer in a tuneable open-access microcavity containing oligonucleotide-based molecular beacons with a donor–acceptor pair of closely located FAM and TAMRA organic dyes exhibiting an efficient FRET outside the microcavity.

In this study, we have measured the dependence of the photoluminescence (PL) properties of the molecular beacon solution in a microcavity and have analysed the dependence of the polaritonic state population on the detuning of the optical microcavity mode. We have used the Jaynes–Cummings model to calculate the eigenfunctions of the strongly coupled three-component system and have estimated the mixing of excitons and photon fractions in hybrid states. We have also estimated the possibility of changing the relaxation pathways by varying the degree of the exciton–photon mixing in polaritonic states. Furthermore, we have explored the particular situation when the donor is much stronger coupled with the optical mode than the acceptor. For the best of our knowledge, we are the first to experimentally demonstrate the possibility to reverse the donor and acceptor roles ("carnival effect") within a donor–acceptor pair of organic dyes, which has been theoretically investigated by the group of Joel Yuen-Zhou [29].

## 2. Results and Discussion

In order to investigate the feasibility of controlling the resonant energy transfer in a donor–acceptor pair of closely located organic dyes, we have employed a tuneable microcavity with a relatively small mode volume previously developed in our group [30,31]. Briefly, the tuneable microcavity unit was composed of plane and convex mirrors that form an unstable λ/2 Fabry–Perot microcavity (**Figure 1a**, see the SI for details). The upper mirror was made convex in order to satisfy the plane-parallelism condition at one point, thus minimizing the mode volume. The plane bottom mirror was mounted on top of a Z-piezo positioner, which provided fine tuning of the microcavity length in the range of up to 10 μm with a nanometre precision. The plane–convex design of the tuneable microcavity is characterized by rather high quality factors (up to several hundred units), whereas the mode volumes can be as low as tens of $\left(\frac{\lambda}{n}\right)^3$, thus combining the advantages of both optical and plasmonic cavities [31]. In this study, the quality factor of the microcavity mode was about 35 and the mode volume was about $15 \times \left(\frac{\lambda}{n}\right)^3$, for all the detunings used. Previously, we have demonstrated the advantages of the developed tuneable microcavity, such as a controllable distance between the mirrors with a nanometre accuracy and a small mode volume, which result in a much higher Rabi splitting energies compared to the standard optical microcavities [28]. In particular, we have demonstrated a strong coupling of the ensemble of Rhodamine 6G molecules with a Rabi splitting as high as 225 meV at room temperature, which has been previously shown only for the case of surface plasmon–polaritons. A drawback of the tuneable setup developed is that the effect of strong coupling cannot be simply observed in the transmission spectra, because the concentration of the molecules inside the cavity is relatively low and the number of photons from the white LED used for measurements of the transmission considerably exceeds the number of generated excitons. However, the developed by us microcavity combining the advantages of optical microcavity (low energy dissipation) and plasmonic cavity (small mode volume) significantly increases the number of materials suitable for operation in a strong coupling regime, thus paving the way to plenty of new practical applications. In this study, we employed this set-up to achieve a high Rabi splitting energy of hybrid states formed by the exciton transitions in donor–acceptor pairs of organic dyes exhibiting the FRET effect and the localized resonant electromagnetic field.

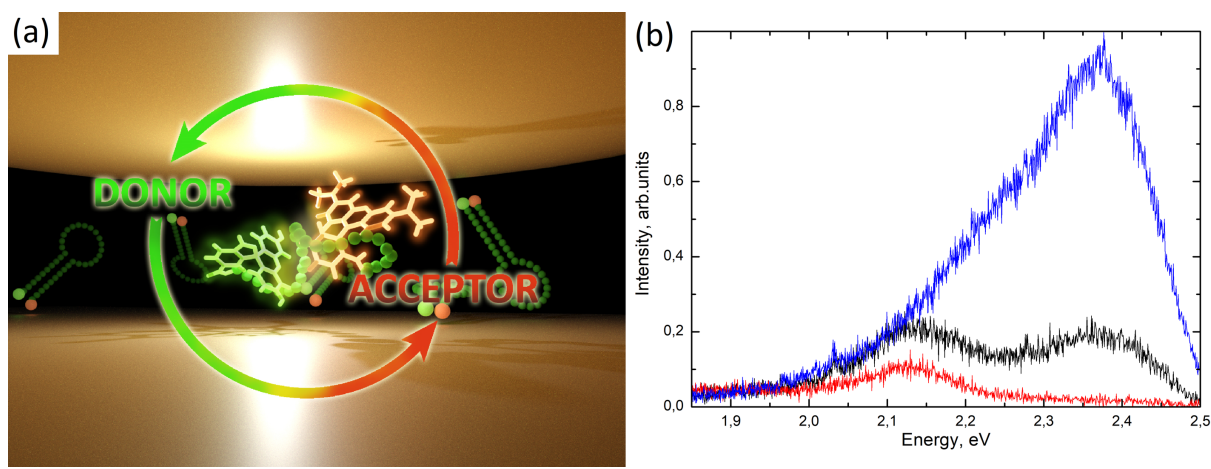

**Figure 1.** (a) The schematic of the tuneable microcavity. (b) The PL emission spectra of the solutions containing molecular beacons labelled with FAM alone (blue), TAMRA alone (red), and both FAM and TAMRA (black) dyes. All samples were located outside the microcavity. The PL spectra were excited at 450 nm. The concentrations of the dye molecules in the sample solutions were about 100 μM in all experiments. The molecular beacons are drawn not to scale.

The donor–acceptor pairs of organic dyes employed in our experiments consisted of 6-carboxyfluorescein (FAM) used as a donor and tetramethylrhodamine (TAMRA) as an acceptor. The distance between the donor and acceptor dye molecules conjugated with the opposite termini of the oligonucleotide-based molecular beacon is determined by the diameter of the DNA double helix (about 2 nm), which is small enough to ensure efficient direct dipole–dipole coupling between them (**Figure S1**) [32]. The details on the chemical structures of the oligonucleotide-based molecular beacon and dye molecules, as well as on the optical properties of the dyes, are presented in the SI.

**Figure 1b** shows the PL spectra of oligonucleotide-based molecular beacons labelled with FAM alone, TAMRA alone, or both FAM and TAMRA dyes placed onto the bottom mirror of the setup that were measured in the absence of the upper mirror; i.e., the beacons were located outside the microcavity.

FAM is a well-known organic dye with a quantum yield as high as 97% [33], with the main exciton transition characterized by a relatively large transition dipole moment ranging from 7 to 12 D [33] and the main emission peak maximum at about 2.36 eV. The TAMRA dye has a much lower quantum yield of about 22% [34], as well as lower values of the transition dipole moment [35]. The PL emission maximum of the TAMRA dye solution is about 2.13 eV. However, it is noteworthy that the transition dipole moments and PL quantum yields of the dyes may be changed upon their conjugation with the oligonucleotide due to the possible appearance of new intermolecular interactions in the conjugated samples. **Figure 1b** shows that the PL spectra of the FAM–TAMRA donor–acceptor pair excited at 450 nm could be obtained by linear superposition of the PL spectra of these dyes measured separately, which indicates the absence of direct ground-state interaction between these dye molecules in the molecular beacon. Despite the low quantum yield, the PL emission from TAMRA was stronger than that from FAM in the case of the donor–acceptor pair operating in the FRET regime. The efficiency of the resonance energy transfer from FAM to TAMRA molecules was estimated to be about 80% (see the SI for details).

We further analysed the PL spectra of the dyes conjugated with the oligonucleotide of the molecular beacon alone and in the form of a donor–acceptor pair operating in the FRET regime and placed into the tuneable microcavity at different cavity detunings (**Figure 2**). The cavity mode tuning has been performed by changing the distance between the microcavity mirrors from 735 to 945 nm at 15-nm steps.

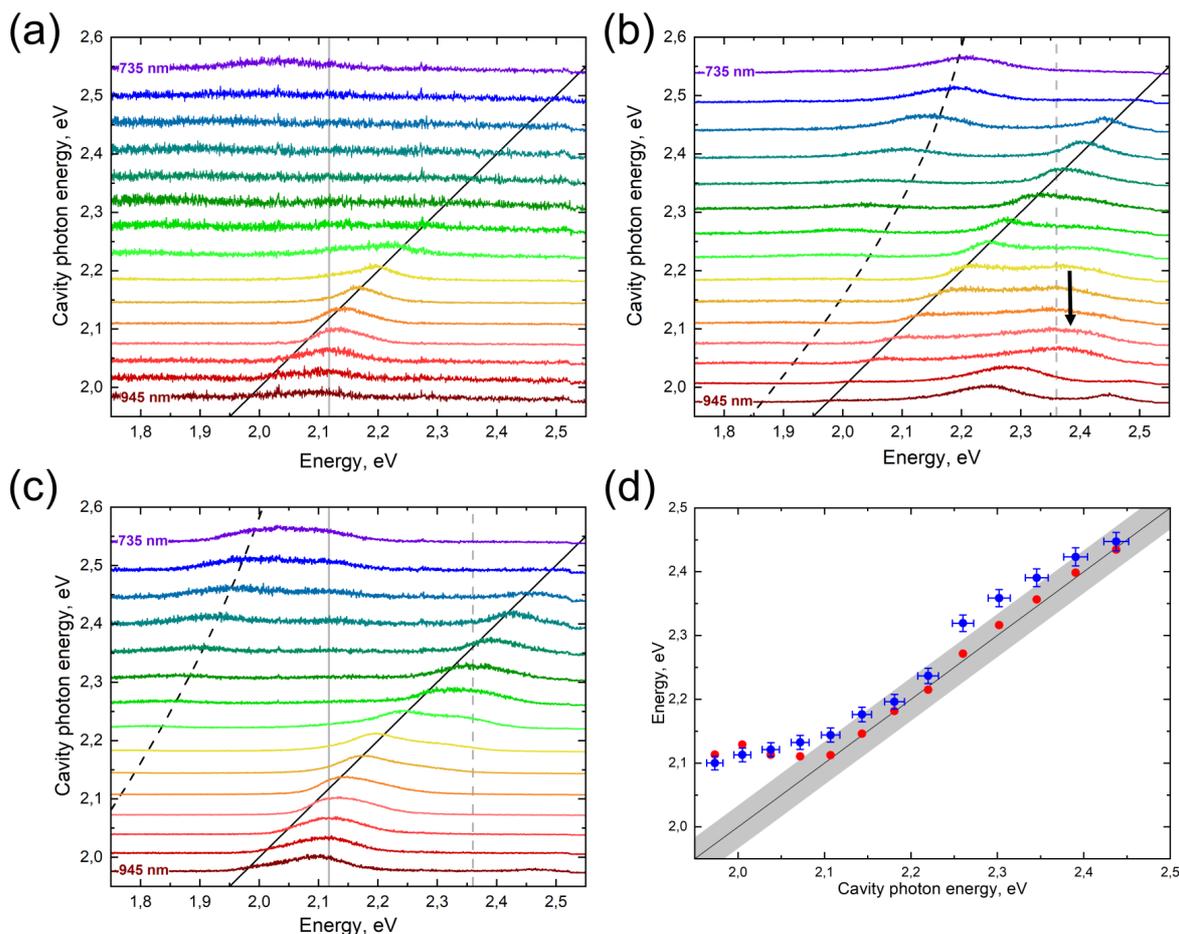

**Figure 2.** The PL spectra recorded at distances between the mirrors varied from 735 to 945 nm at a 15-nm step for microcavity filled with the molecular beacons labelled with TAMRA (a), FAM (b), or the FAM–TAMRA donor–acceptor pair operating in the FRET regime (c). The black solid line shows the position of the cavity mode; the vertical grey solid and dashed lines show the positions of the acceptor and donor excitons, respectively; the black dashed lines mark the calculated lower polaritonic branches. (d) The experimental (blue dots) and calculated (red dots) spectral positions of the - peak corresponding to the emission from weakly coupled states. Black arrow shows the approximate position of the emission peak corresponding to the upper polariton (b).

The corresponding PL emission spectra are shown in **Figures 2a–2c**. It is worth mentioning that the specific properties of our microcavity allowed us to detect the PL emission from molecules weakly coupled to the transverse modes of both the lowest and higher orders of the cavity (see the SI for details). This PL associated with the enhanced emission peak following the cavity mode, slightly blue-shifted relative to the cavity photon spectral position corresponding to the lowest-order transverse mode of the cavity. Qualitatively, this blue shift can be understood considering the difference in signal collection efficiency between the transmission and PL measurements (see the SI for details). The resulting emission corresponding to this peak is satisfactorily fitted in the weak coupling approximation. **Figure 2d** presents the experimentally measured dependence of the spectral position of the emission maximum associated with the weakly coupled molecules on the cavity mode detuning extracted from **Figure 2c** and the corresponding model calculations. It can be seen that the calculated dependence is in good agreement with the experimental data, the difference not exceeding the full-width-at-half-maximum (FWHM) of the cavity mode. It is also important to note that the large shift of the emission peak from the cavity mode in the region of low mode energies was due to the poor spectral overlap of the dye PL spectra with the lowest-order transverse cavity mode. Similar PL emissions from weakly coupled molecules were observed for the FAM and TAMRA dyes separately (**Figures 2a, 2b**). It is noteworthy that, due to the non-resonant excitation through the lower mirror of the microcavity, the power of excitation inside the cavity depended on the cavity detuning. Although the intensities of the measured PL spectra presented in **Figure 2** are not calibrated against the excitation power, the necessary corrections have been made in the analysis presented below.

The molecular beacons labelled with the TAMRA dye alone and placed into the microcavity exhibited only the emission from weakly coupled molecules (**Figure 2a**), which can be explained by higher losses, lower transition dipole moments of the rhodamine derivative, and the resultant low coupling strength [35]. However, it is noteworthy that the change in the emission intensity from the weakly coupled states of TAMRA was much bigger than that for the case of the molecular beacon labelled with the FAM dye alone. Indeed, the Purcell PL intensity enhancement for emitters with a lower quantum yield is known to be stronger than that for emitters with a higher quantum yield, because the Purcell effect changes only the radiative relaxation rate [36].

The PL of molecular beacons labelled with the FAM dye alone and placed into the microcavity is shown in **Figure 2b**. Considering the impact of the ensemble of weakly coupled molecules, one can clearly see the anticrossing behaviour of the PL emission peaks for cavity energies larger than 2.25 eV. The positions of the low-energy peaks in the emission spectra corresponding to the lower polaritonic branch (LPB) allow one to estimate the Rabi splitting energy to be about 457 meV, which is a rather high value for this type of a cavity. It is well known that the observation of the emission from the upper polariton branch (UPB) is difficult due to the fast relaxation of the UPB to the exciton reservoir [37]. In the PL experiments we have observed the emission peak that could be associated with the upper polariton for deeply negative detuning, which is marked in **Figure 2b**. However, it is difficult to estimate the exact positions of the associated emission spectra for most other values of detuning due to the emission from bare donor molecules, the large spectral broadening, and the limited spectral range of detection in the setup used.

**Figure 2c** shows the PL spectra recorded for different detunings of the microcavity containing molecular beacons labelled with both FAM and TAMRA dyes. As in the case of the molecular beacons labelled with the FAM or TAMRA dye alone, one can see a peak corresponding to weakly coupled molecules, whose spectral position follows the cavity photon energy with a small blue shift. The experimental and calculated dependences of the spectral position of this peak on the cavity photon energy are shown in **Figure 2d**. It is worth mentioning that the relative intensity of this peak was changed relative to the emission peak of the beacons labelled with FAM alone due to the resonance energy transfer between bare donor and acceptor states. Indeed, its intensity was lower than that in the case of molecular beacons labelled with FAM alone in the donor spectral region and higher than that in the case of molecular beacons labelled with TAMRA alone in the acceptor spectral region. These effects were certainly due to the remaining highly efficient resonance energy exchange through dipole–dipole interaction upon the deposition of the molecular beacons labelled with both FAM and TAMRA dyes operating in the FRET regime inside the cavity. It is important to note that, in contrast to the case of molecular beacons labelled with the FAM dye alone, no signal from bare donor states was detected at negative detunings in the cavity photon energy range from 2.0 to 2.3 eV. The fact that the donor–acceptor dye pair outside the cavity exhibited emission from both donor and acceptor dyes with a FRET efficiency of about 80% indicated that the ratio between the donor–acceptor energy transfer rate and the donor radiative relaxation rate inside the cavity was increased. This can be explained by taking into account the suppression of the spontaneous emission of bare donor states when the cavity mode was shifted from the donor emission bandwidth. The low density of photonic states beyond the cavity mode compared to the density in the free space led to a decrease in the spontaneous emission rate and, consequently, to the effective increase in the FRET efficiency.

However, at positive detunings, we observed a PL peak determined by emission from the LPB, as in the case of the cavity containing molecular beacons labelled with the donor dye alone, which was red-shifted relative to the emission spectrum of the uncoupled dye molecules. Similarly to the previous cases, we were unable to detect any emission that could have been attributed to the UPB due to the typically fast energy relaxation from the upper polariton to the donor exciton reservoir. Thus, we omit the UPB from most of the discussion. Nevertheless, the presence of strong coupling was clearly evidenced by anticrossing of the LPB at the cavity detunings where the cavity photon mode and donor excitons had to be degenerated. Considering the exitonic constituents as independent harmonic oscillators, we assumed that our cavity could couple together three oscillators, FAM, TAMRA, and the cavity mode. Therefore, we characterized the observed dispersion by three polariton branches. In the case of a negatively detuned cavity, we observed an emission peak that could be attributed to the middle polariton branch (MPB). However, due to the pronounced emission from weakly coupled states and large broadening of the emission spectra, this part of the emission spectrum can hardly be used to quantitatively analyse the spectral position of the MPB.

The dependence of the polariton branches on the cavity photon energy presented in **Figure 3** confirms that strong coupling occurred in this case. Experimental data were derived from the peak energies in the PL spectra for different distances between the cavity mirrors shown in **Figure 2c**. In order to obtain a satisfactory fitting, we used the Jaynes–Cummings model assuming coupling between the cavity mode and the donor and the acceptor excitonic states resulting in the formation of three polaritonic branches and varied the strength of coupling between the cavity photon and both excitonic transitions (see the SI for details about the fitting). The best fit was achieved for the strengths of coupling with the cavity mode of 41 and 435 meV for the acceptor and donor excitons, respectively.

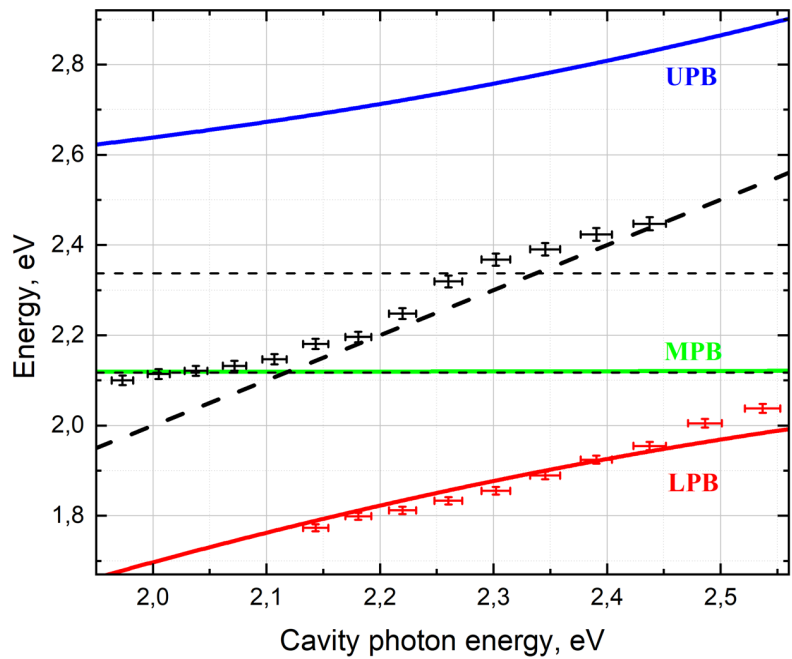

**Figure 3.** The energies of the lower (red), middle (green), and upper (dark blue) polariton branches at different cavity detunings experimentally derived from the PL spectra (dots) and theoretically calculated (solid lines). The black dots correspond to the emission from weakly coupled dye molecules. The horizontal and inclined black dashed lines show the positions of the donor and acceptor excitons and the cavity mode energy, respectively.

The considerably higher value of coupling strength obtained for the donor molecule compared to the acceptor one is explained by the larger transition dipole moment of this dye. **Figure 3** also shows the energy of an uncoupled cavity photon, exciton transitions for uncoupled donor and acceptor dyes, and the energies of the emission maxima for the weakly coupled part of the molecular ensemble.

In order to determine the Hopfield coefficients, that describe the donor exciton, acceptor exciton, and photon mixing for the given coupling strengths, we further calculated the eigenfunctions of the Jaynes–Cummings Hamiltonian and represented each of them as a superposition of the initial pure photon and exciton states (see the SI for details). **Figure 4** shows the Hopfield coefficients for each polariton branch and their dependences on the cavity detuning. In the analysed PL spectral region, the polariton branches displayed quite different behaviours.

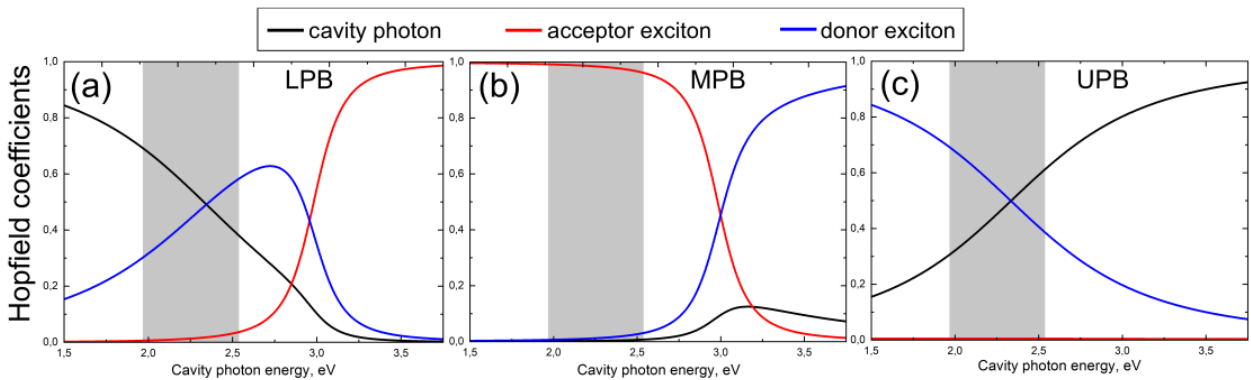

**Figure 4.** The Hopfield coefficients of the lower (a), middle (b), and upper (c) polaritonic branches calculated using the Jaynes–Cummings model for the microcavity filled with molecular beacons labeled with both FAM and TAMRA dyes. The grey area shows the range of the cavity mode detuning measured experimentally.

It can be seen from **Figure 4** that the upper and the lower polaritons mainly consist of the donor exciton and photon fractions, whose ratio is reversed for both branches upon cavity tuning. On the other hand, for the MPB, the contribution from the acceptor exciton strongly dominates the other components. However, this contribution decreases with an increase in the cavity mode energy in the same way as the relative contribution of the acceptor exciton increases significantly in the LPB. Such a redistribution of fractions in the polariton branches is in accordance with the exciton–photon coupling strengths, which differ significantly for the donor and acceptor excitons, as we have mentioned above.

In order to quantitatively correlate the polariton population dependence on the detuning and the PL intensity measured experimentally, one need to introduce corrections arising from variations of both the photon fraction and the excitation intensity [25,26]. The finite elements method was used to calculate the dependence of the excitation field intensity inside the cavity on the detuning (see the SI for details). In order to make the necessary corrections, we divided the measured PL intensity of the LPB by the integral intensity of non-resonant excitation at 450 nm over the volume of the microcavity in the region of cavity detuning. Finally, the relative polariton population (**Figure 5**) was obtained using the following equation [25]:

$$N_{LPB}(\hbar\omega_{cav}) \propto \frac{I_{LPB}(\hbar\omega_{cav})}{C^{cav}_{LPB}(\hbar\omega_{cav})}, \tag{1}$$

where $I_{LPB}$ is the experimentally observed intensity of PL from the LPB corrected for the excitation intensity, and $C^{cav}_{LPB}$ is the Hopfield coefficient for the LPB defining the fraction of the cavity photon. The intensity of emission from the polariton state depends linearly on the cavity photon fraction.

As can be seen from **Figure 5**, the population of the LPB non-linearly increases with increasing cavity photon energy. According to the Hopfield coefficient distributions, this growth is accompanied by a linear decrease in the donor exciton fraction approximately from 0.3 to 0.6 and by a drastic increase in the acceptor exciton fraction in the LPB by almost an order of magnitude.

Now, we will analyse the population and depopulation mechanisms for the LPB using an approach described in detail in Ref. [25]. The simplified consideration is based on the assumptions on fast relaxation from the UPB to the donor excitonic reservoir (which typically occurs on the femtosecond timescale [37]) and efficient FRET of most of the energy to the acceptor reservoir. Therefore, the number of states in the acceptor excitonic reservoir can be considered constant. In principle, there are three different mechanisms determining the LPB population: scattering with vibrations from both excitonic reservoirs of the donor and acceptor and direct radiative pumping. The efficiencies of the first two mechanisms strongly depend on the corresponding exciton fraction in the LPB and should be changed with the detuning. The radiative pumping mechanism is a direct absorption of the photon emitted by the weakly coupled exciton transitions, which is accounted for by the photonic fraction in the LPB. However, it is almost negligible for cavities with a short cavity photon lifetime that contain an ensemble of molecules with a low optical density. The depopulation of the LPB occurs via radiative and non-radiative relaxations, which depend on the photon and exciton lifetimes, respectively. Thus, the mean polariton population ($N_{LPB}$) in the steady state $\left(\frac{dN_{LPB}}{dt}=0\right)$ can be described by the following equation:

$$\frac{dN_{LPB}}{dt} = A_1 C_{LPB}^A (B_A+1) + A_2 C_{LPB}^D (B_D+1) + A_3 C_{LPB}^{cav} - A_4 N_{LPB} C_{LPB}^{cav} - A_5 N_l \quad (2)$$

where $A_{1,2,3}$ are the proportionality constants for the terms corresponding to the LPB population through vibration scattering from the acceptor and donor reservoirs and direct radiative pumping, respectively; $A_{4,5}$ are the proportionality constants for the terms describing the depopulation via radiative ($A_4$) and non-radiative ($A_5$) relaxations. The first two terms describe the processes accompanied by the emission of molecular vibrations, which depend on the Bose–Einstein distribution $B_j = \left(\exp\left|\frac{E_j - E_{LPB}}{kT}\right| - 1\right)^{-1}$ ($j = A, D$), where $E_j$ is the bare exciton energy, $E_{LPB}$ is the energy of the LPB, $T$ is the temperature in kelvins, and $k$ is the Boltzmann constant. For the approximation of the population of the LPB in the steady state (**Figure 5**), we used the following equation, which could be easily obtained from Equation (2) under some reasonable assumptions described below:

$$N_{LPB} = \frac{A_1 C_{LPB}^A (B_A + 1) + A_2 C_{LPB}^D (B_D + 1) + A_3 C_{LPB}^{cav}}{A_4 C_{LPB}^{cav} + A_5} \approx \frac{A_1 C_{LPB}^A + A_2 C_{LPB}^D}{A_4 C_{LPB}^{cav}} \quad (3)$$

To derive Equation (3), we first assumed that the radiative pumping of the LPB is negligible compared to the vibrational scattering due to the low optical density of the medium inside the cavity and low cavity Q-factor. Second, non-radiative relaxation of LPB to the exciton reservoirs strongly depends on the local vibrational environment and assumed to be negligible comparing to the radiative decay mechanism [25,38], which is determined by the cavity photon lifetime (less than 10 fs). Thus, radiative decay through the photonic fraction becomes a prevailing depopulation mechanism, resulting in $A_4 \gg A_5$. Finally, our experimental conditions $(B_{A,B} + 1) \approx 1$ should be taken into consideration. In order to obtain the best fit, we minimized the standard deviation by varying the ratio between the parameters $A_{1,2,4}$. The best fit obtained with the use of this model and the experimentally observed relative population of the LPB are shown in **Figure 5**.

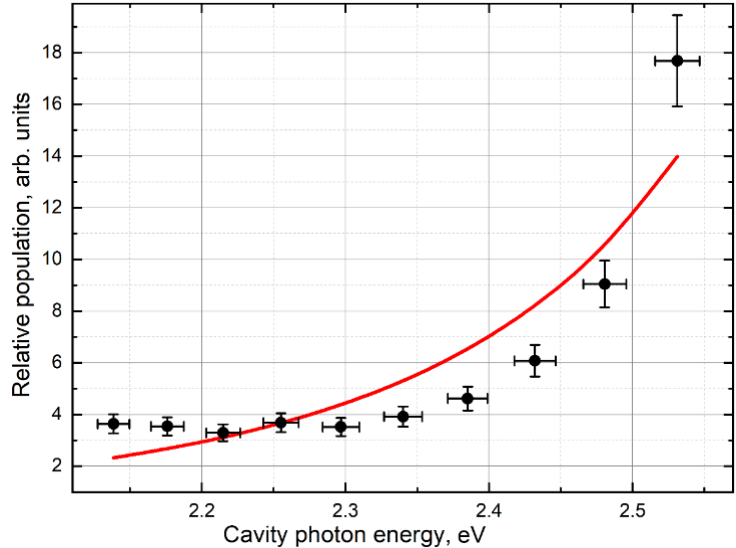

**Figure 5.** Experimentally observed data (black dots) and theoretically modelled results (red line) for dependences of the relative population of the lower polaritonic branch on the cavity detuning.

It is important to note that the best fit was obtained with $A_2$ tending to be zero. This corresponds to the low efficiency of the LPB population caused by scattering from the donor exciton reservoir, which may have been due to the relatively low rate of this process compared to the depopulation of the donor excitonic reservoir through FRET to the bare acceptor states. This mechanism was previously shown to be dominant over polariton-assisted energy transfer in mixed donor–acceptor ensembles [26]. Thus, the energy relaxation pathways in our system can be described as follows (**Figure 6**).

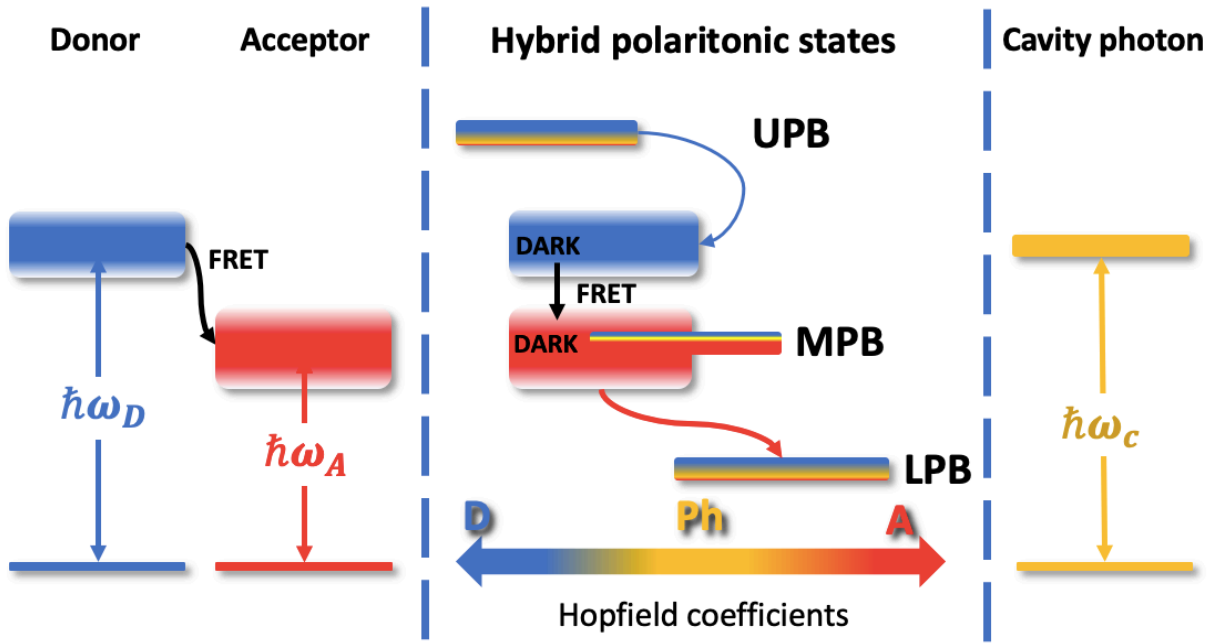

**Figure 6.** Schematic energy-level diagram showing the energy relaxation pathways in the system with predominant strong coupling to the donor excitonic transition. The thickness of the horizontal lines reflects the density of the bare donor and acceptor excitonic states, as well as three new hybrid light–matter states: the upper (UPB), middle (MPB), and lower (LPB) polariton branches.

The non-radiatively pumped population of the UPB rapidly decays to the donor exciton reservoir [37]. Then, due to the short distance between the donor and acceptor dye molecules in the molecular beacon, direct dipole–dipole FRET occurs with the efficiency close to unity. The FRET efficiency for the molecular beacon placed into the microcavity was found to be increased compared to that for the molecular beacon outside the cavity (about 80%), because we did not observe any emission from the bare donor states at negative detuning, in contrast to the donor-only case. This may have been due to the decreased rate of radiative relaxation of bare states at negative detunings, leading to the increase in FRET efficiency. The most interesting is that, once the energy had been transferred to the acceptor excitonic reservoir, it started to populate the LPB, which is mostly a mixture of the donor and cavity photon fractions due to the much higher coupling strength between the donor and the cavity photon. Thus, vibration scattering from the acceptor reservoir was shown to be the main population mechanism of the lower polariton state with the donor exciton fraction exceeding the acceptor one. It was demonstrated previously that a small absolute value of the specific exciton fraction in the polariton branch still allowed energy transfer with the corresponding excitonic reservoir [25]. In our experiments, the population of the LPB depended on the relative variation of a small fraction of the acceptor exciton in the polariton state despite the considerably higher absolute value of the Hopfield coefficient, corresponding to the donor exciton. Finally, we can state that we have engineered a strongly coupled system with donor–acceptor role reversal or the "carnival effect" [27]. Indeed, we have developed a system with dominant coupling between the donor and acceptor leading to the formation of a donor-like polariton state with the lowest energy in the system. This allowed energy transfer first from the donor exciton reservoir to the acceptor exciton reservoir via standard FRET and then from the acceptor reservoir to the donor-like lower polariton state.

## 3. Conclusions

We have investigated strong coupling between the optical modes of a tuneable microcavity and the excitonic transitions of two closely located organic dye molecule labels of an oligonucleotide-based molecular beacon. The anticrossing dependence of the emission spectra of the donor–acceptor dye pair operating in the FRET regime on the detuning of the microcavity mode has been demonstrated by varying the distance between the cavity mirrors. We have estimated the dependence of the polaritonic state population on the detuning and photon–exciton mixing, which has been calculated by fitting the experimental results with the three-level Jaynes–Cummings model and rate equations. In addition to the efficient FRET between the bare exciton states, significant alteration of the relaxation pathways by changing the photon and exciton mixing in the lower polariton state has been demonstrated.

We have confirmed that, in our system of molecular beacons labelled with organic dyes and located inside a tuneable open-access microcavity, the resonance energy transfer via direct FRET remains the dominant process despite the strong coupling of the dye excitons to the cavity mode, as has been previously demonstrated [26]. However, we have shown that the PL emission of the systems with strong dipole–dipole interaction can nevertheless be altered by strong coupling of their exciton transitions to the cavity photon. First, we have demonstrated a significant increase in the efficiency of energy transfer from the donor to the acceptor exciton reservoir, which tends to be unity inside the microcavity. Second, despite the efficient energy transfer between the donor and the acceptor exciton reservoir, we have observed emission from the LPB, which has been shown to have a considerably higher donor fraction compared to the acceptor one. Furthermore, we have proved that the LPB is populated mainly through scattering from the acceptor reservoir, despite the much larger absolute value of the donor exciton fraction than the acceptor fraction in the lower polariton. Thus, by obtaining strong coupling of the photonic mode with the exciton transition predominantly for the donor, we have demonstrated the so-called "carnival effect", where the donor and acceptor reverse their roles [29]. Consequently, energy transfer occurs first from the donor to the acceptor exciton by means of resonant dipole–dipole interaction and then from the excitonic states in the acceptor reservoir to the mainly-donor LPB. We speculate that our experimental findings, together with further investigations of the interplay between the polaritonic states and excitonic reservoirs, as well as control of the relaxation pathways, will pave the way to new applications of strong coupling to remote-controlled chemistry, energy harvesting, energy transfer and sensing.

**Experimental Section**

***Sample preparation.*** All components of the oligonucleotide-based molecular beacon labelled with donor or acceptor dye alone or both dyes operating in the FRET regime were obtained from Evrogen (Evrogen Joint Stock Company, Moscow, Russia) and diluted in phosphate-buffered saline (PBS) to obtain a concentration of 100 μM. Detailed information on the chemical structures and optical properties of the dyes and all three types of molecular beacons used are presented in the SI. For experiments, 10 μL of each sample was placed on the flat bottom mirror of the microcavity and then covered with the convex upper mirror. During the experiments, the surface exposed to air was small enough to ensure a negligible rate of sample solvent evaporation and a constant sample concentration.

***The tuneable microcavity.*** The sketch of the experimental setup consisting of a tuneable microcavity and an optical excitation/collection system is shown in the SI. The setup is described in detail elsewhere [30,31]. The plane-parallelism point and the sample were aligned by moving the convex mirror in the lateral direction by means of an XY precision positioner. The curvature radius of the upper convex mirror was 77.3 mm. The sample was deposited directly onto the plane mirror, which consisted of standard (18×18 mm) glass coverslips covered with a ~35-nm layer of aluminium and a 20-nm protective layer of $SiO_2$.

***Transmission and PL measurements.*** The tuneable microcavity as a part of the unique setup called the *System for Probe-Optical 3D Correlative Microscopy* (http://ckp-rf.ru/usu/486825/) was mounted on an inverted confocal microspectrometer consisting of an Ntegra-base (NT-MDT) with a 100X/0.80 MPLAPON lens (Olympus) on a Z-piezo positioner, an XY scanning piezo stage, and a homemade confocal unit. The MCWHF2 white LED (Thorlabs) with an optical condenser was used for transmission measurements. In PL measurements, we used emission of a 450-nm L450P1600MM laser (Thorlabs) with an LDS5-EC (Thorlabs) power supply, with a pump power of about 2.1 mW for excitation. The pump power in all the PL experiments was far from the saturation threshold. The recording system included an Andor Shamrock 750 monochromator equipped with an Andor DU971P-BV CCD (Andor

Technology Ltd.) camera and two 488-nm RazorEdge® ultrasteep long-pass edge filters (Semrock).

**Supporting Information**
The Supporting Information is available from the Wiley Online Library or from the author. It includes: detailed descriptions of the properties of the molecular beacon samples, tuneable microcavity setup, and PL/Transmission collection system; calculations of the pumping intensity dependence on the cavity detuning; and description of the Jaynes–Cummings model.

**Acknowledgements**
This study was supported by the Ministry of Education and Science of the Russian Federation (grant no. 14.Y26.31.0011). I.N. acknowledges the support from the Ministry of Higher Education, Research and Innovation of the French Republic and the University of Reims Champagne-Ardenne. Y.R. acknowledges the support from the Basque Government (grant no. IT1164-19). We thank Vladimir Ushakov for the help with technical preparation of the manuscript.

## S1 Materials

6-Carboxyfluorescein (FAM) was selected as donor molecule, Carboxytetramethylrhodamine (TAMRA) as acceptor molecule. The most common way to fix the required for efficient FRET distance between chromophores is to use oligonucleotide-based molecular beacons as in this case distance between donor and acceptor is of the order of the diameter of the DNA double helix, which is 2 nm.

In this study, donor and acceptor were conjugated with self-complementary oligonucleotide, 5'-TGG AGC GTG GGG ACG GCA AGC AGC GAA CTC AGT ACA ACA TGC CGT CCC CAC GCT CCA-3'.

Donor- and acceptor-only labeled hairpins were also obtained and studied as controls. (**Figure S1**). Oligonucleotide sequence of 57 units with 18 base pairs was selected so that to ensure hairpin stability and small distance between donor and acceptor. Molecular weights of donor-only, acceptor-only and both donor and acceptor labeled hairpins are $M_{FAM}$=17477.47, $M_{TAMRA}$=17568.68 and $M_{FAM\_TAMRA}$=17945, respectively. All components were obtained from Evrogen (Evrogen Joint Stock Company, Moscow, Russia) and diluted in phosphate buffered saline (PBS) to obtain concentration of 100 μM. For experiments, we used 10 μL of each component. Photoluminescence and absorption spectra of the compounds were first studied outside of the cavity (**Figure S2**). Here we used different excitation wavelength depending on the absorption spectra of each compound.

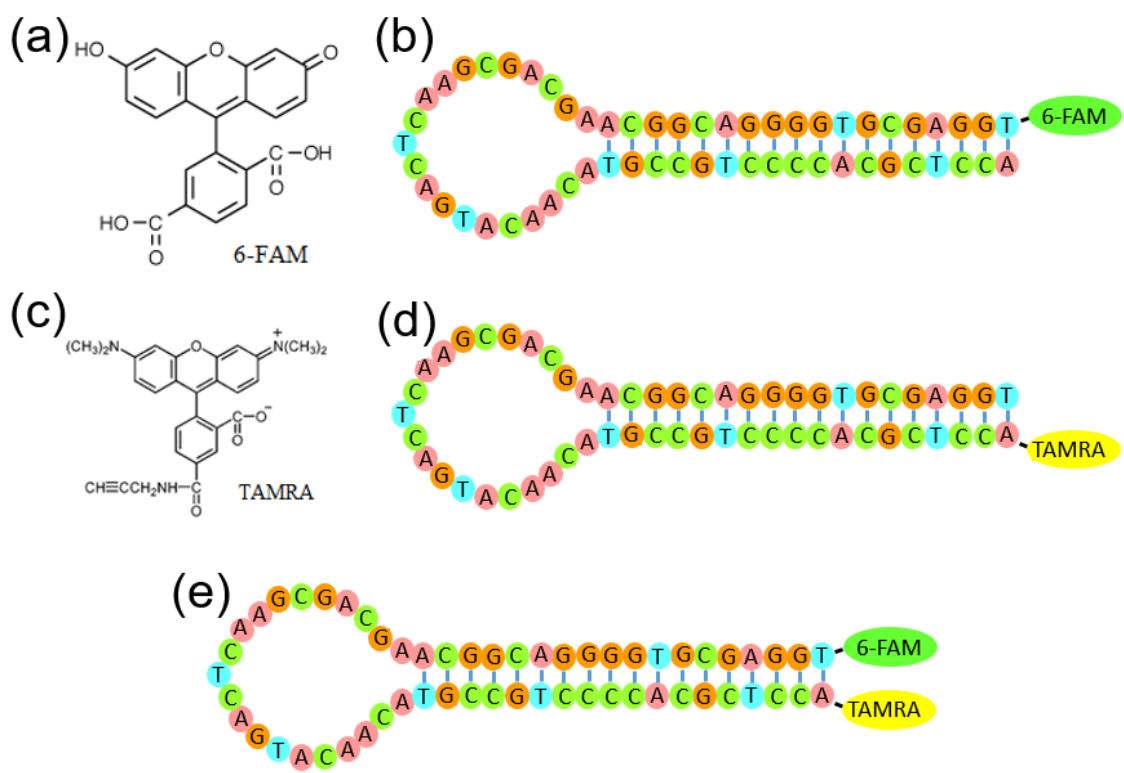

**Figure S1.** Structures under study. Donor (a) and acceptor (c) chemical structures; schematic for the structure of molecular beacons: donor-only labeled hairpin (b), acceptor-only labeled hairpin (d), both donor and acceptor labeled hairpin (e).

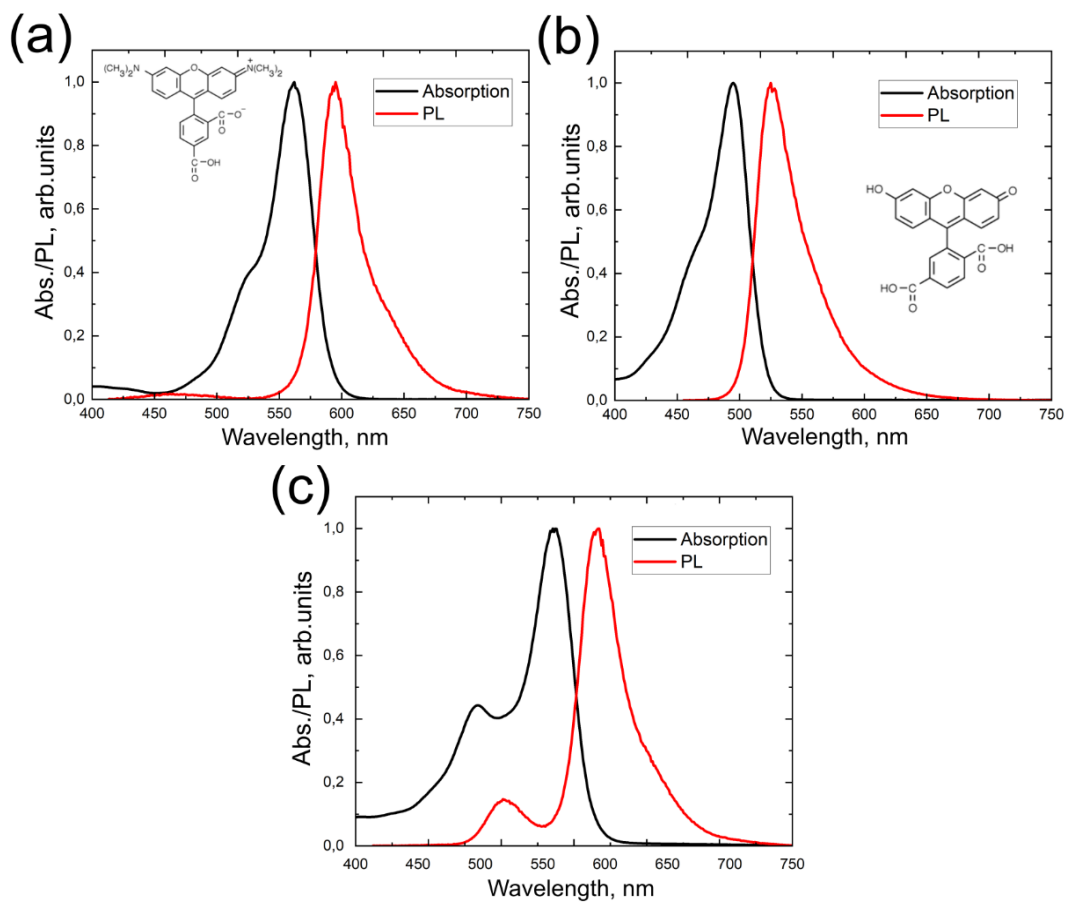

**Figure S2.** Absorption and photoluminescence spectra of solutions of the TAMRA dye (a); FAM dye (b); FAM and TAMRA dyes (c).

In order to estimate FRET efficiency via dipole-dipole interaction we placed solution containing beacons labeled with FAM, TAMRA, and FRET pair alternately on the lower mirror of the microcavity without the upper one and excited it non-resonantly with 450 nm laser. In such way we ensured the same experimental conditions as in experiments with the media placed inside the cavity. The measured Pl spectra are shown in **Figure S3**.

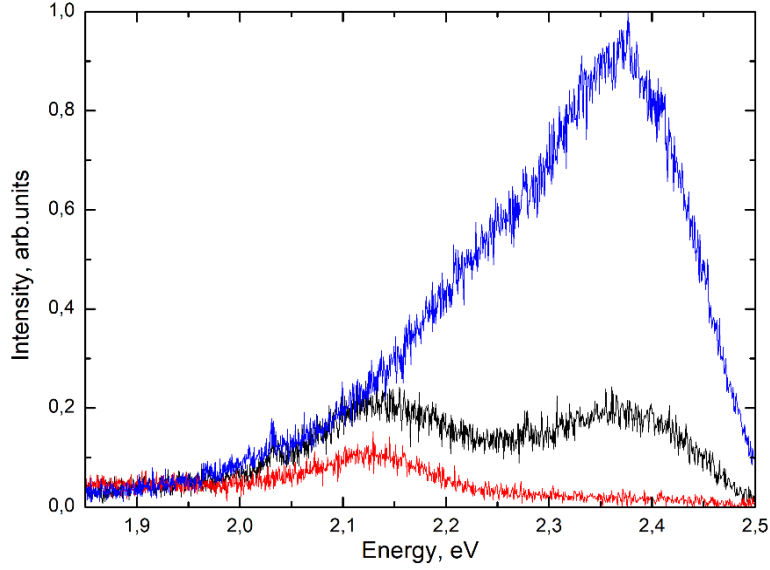

**Figure S3.** PL emission spectra of the solutions containing molecular beacons labelled with FAM (blue), TAMRA (red), FAM and TAMRA (black) outside of the cavity.

From these spectra it can be seen that FAM-only labeled hairpin shows high fluorescence (probably due to FAM high quantum yield of 97%) intensity with a peak at 525nm while TAMRA-only counterpart shows low intensity. These results correspond to higher efficiency excitation at 450 nm of FAM ( ~20% of maximum at 495 nm) compared to TAMRA (~2% of maximum at 546 nm). Both donor and acceptor labeled hairpin shows significant decrease in donor intensity and simultaneous increase in acceptor intensity, a result that indicates energy transfer from donor to acceptor. From these data FRET efficiency can be estimated using following expression [s1,s2].

$$E = 1 - \frac{F_{DA}}{F_D}, \qquad (S1)$$

where $E$ is the efficiency of FRET, $F_{DA}$ and $F_D$ are the donor fluorescence in the presence and absence of acceptor, respectively. These spectra lead to estimated FRET efficiency value of 80%.

## S2 Experimental setup

Experimental setup is shown in **Figure S4**. In order to enter strong coupling regime, tunable microcavity, originally introduced in [s3], was used. Briefly, our versatile tunable microcavity cell (VTMC) [s4] is composed of plane and convex mirror that form unstable λ/2 Fabry–Perot microcavity. One mirror is made convex in order to satisfy the plane-parallelism condition and minimize the mode volume. Plane mirror is mounted on top of a Z-piezopositioner to provide fine tuning of a microcavity length in a range up to 10 μm with a nanometer precision, while landing procedure is carried out by the high precision differential micrometer DRV3 (Thorlabs) which is indirectly connected to the convex mirror. The alignment of the plane-parallelism point and a sample is made by moving convex mirror in lateral direction by means of the XY precision positioner. A sample is deposited directly onto the plane mirror that consists of standard (18x18 mm) glass coverslips with ca 35 nm layer of aluminum metallization (Al) on their upper side. The VTMC is mounted onto inverted confocal microspectrometer consisting of an Ntegra-base (NT-MDT) with a 100X/0.80 MPLAPON lens (Olympus) on a Z-piezo-positioner, an XY scanning piezo-stage and a homemade confocal unit. The fluorescence spectra of each sample were excited by a 2.1 W 450-nm laser (L450P1600MM (Thorlabs) with LDS5-EC (Thorlabs) power supply), while for transmission spectra the MCWHF2 white LED (Thorlabs) with homemade optical condenser was used. It should be noted that in experiments laser power was far from saturation. The registration system includes an Andor Shamrock 750 monochromator equipped with an Andor DU971P-BV CCD (Andor Technology Ltd) and two 488-nm RazorEdge® ultrasteep long-pass edge filters (Semrock).

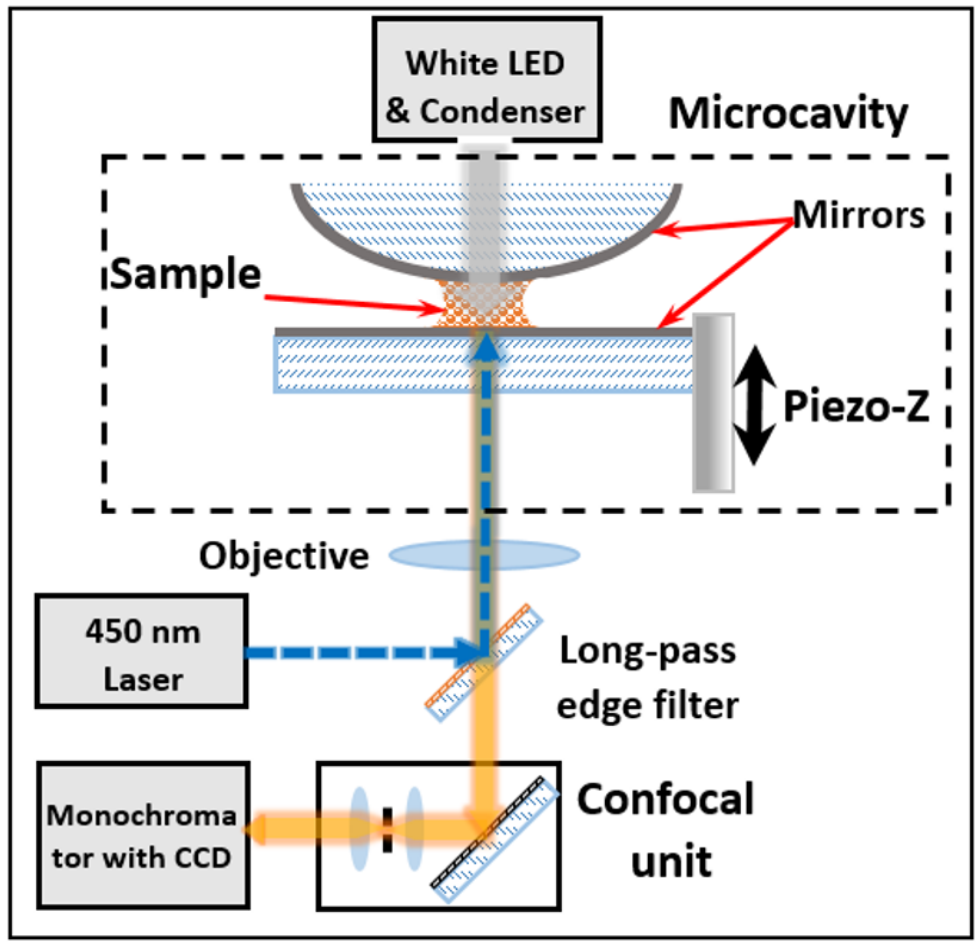

**Figure S4.** Experimental setup

**S3 Accounting for changes in pumping intensity**

For correct calculation of the lower polariton branch population it was necessary to account for changes in pumping intensity during tuning the cavity length. When the exciting field is in resonance with one of the cavity eigenmodes, significant rise in field intensity inside the cavity appears. To account this effect we used numerical model [s5] implemented for calculating spectral and spatial properties of the microcavity electromagnetic modes with the finite elements method [s6]. It should be noted that higher transverse modes of the microcavity were also taken into account. It was necessary because during the excitation process light of the pumping laser was focused by the objective lens with quite high numerical aperture (NA=0.95) and subsequently pumping radiation excited the higher transverse modes (**Figure S5a**). In the transmission experiments illuminating light had approximately planar wavefront (**Figure S5b**), and so there are no appearances of higher transverse modes on the transmission spectra (**Figure S6a**).

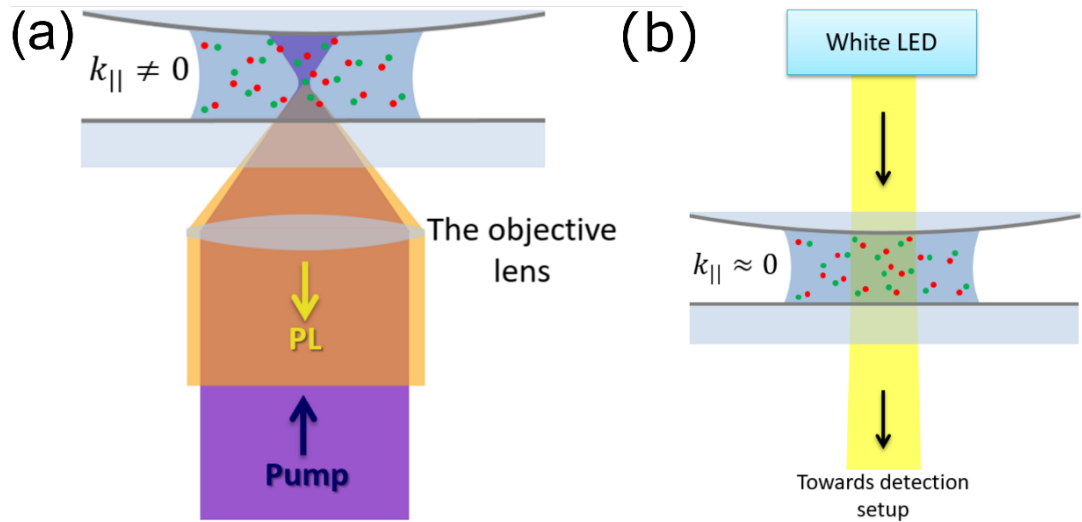

**Figure S5.** The principal configuration of a) the excitation and b) the transmission experiments.

Using developed model, the spectral distribution of the electromagnetic energy was calculated for experimental set of cavity lengths. Then for all this set of spectra the points corresponding to the excited laser frequency were picked. These values were used as the pumping intensity at certain cavity length (**Figure S6b**).

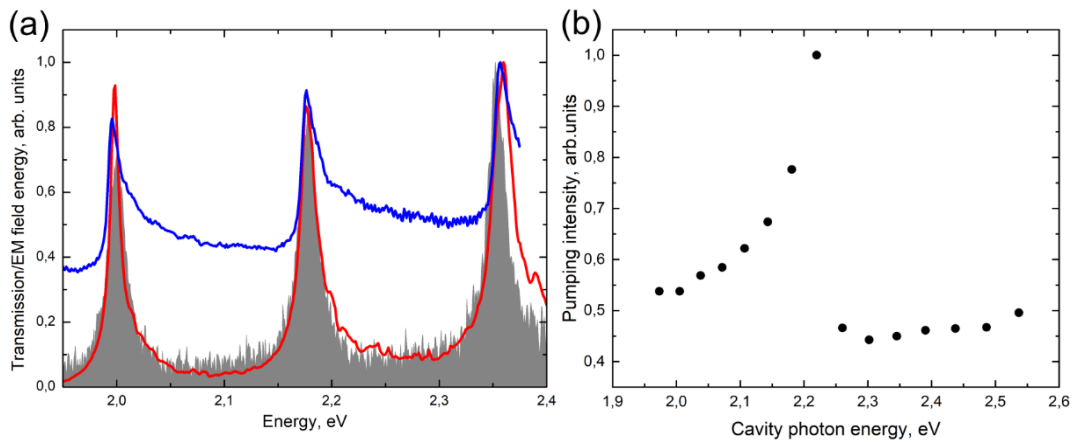

**Figure S6.** Panel (a) shows calculated spectrum of electromagnetic energy inside the microcavity with (red) and without mode selection (blue). The grey area represents corresponding experimental transmission spectrum. Panel (b) shows the pumping intensity dependence on the cavity mode frequency.

## S4 Calculation of the coupling strengths and Hopfield coefficients

Spectral properties of the experimental system based on donor-acceptor pair placed inside the microcavity were researched using the Jaynes-Cummings Hamiltonian that describes interaction between a cavity mode and dipole moments of emitters. For our hybrid system this Hamiltonian is as follows:

$$H_{JC} = \hbar\omega_{cav}a^+a + \frac{1}{2}\hbar\omega_D\sigma_D^z + \frac{1}{2}\hbar\omega_A\sigma_A^z + \hbar g_{D-cav}(a^+\sigma_D^- + a\sigma_D^+) + \hbar g_{A-cav}(a^+\sigma_A^- + a\sigma_A^+), \qquad (S2)$$

where $\hbar\omega_c$, $\hbar\omega_D$ and $\hbar\omega_A$ – energies of microcavity electromagnetic mode, donor and acceptor excitons, respectively; $a(a^+)$- photon annihilation (creation) operator, $g_{j-cav}(j = A, D) = d_j\sqrt{\hbar\omega_c/2\varepsilon_0 V}$ - coupling strength, where $d_D$ and $d_A$ - dipole moments of energy transition of donor and acceptor, $\varepsilon_0$ – vacuum permittivity, $V$ – volume of electromagnetic mode; $\sigma_i^z = |e_i\rangle\langle e_i| - |g_i\rangle\langle g_i|$, $\sigma_i^+ = |e_i\rangle\langle g_i|$, $\sigma_i^- = |g_i\rangle\langle e_i|$ $(i = A, D)$, $|g_i\rangle$ and $|e_i\rangle$ – ground and excited state wavefunctions of emitters. It should be noted here that expression (S1) doesn't have the term describing direct interaction between dipole moments of chromophores. This is due to this process is much slower than energy exchange between emitters and cavity mode for coupling strengths taking place in the current research [s7,s8]. The Hamiltonian also can be written in matrix representation:

$$H_{JC} = \begin{pmatrix} \hbar\omega_{cav} & \hbar g_{D-cav} & \hbar g_{A-cav} \\ \hbar g_{D-cav} & \hbar\omega_D & 0 \\ \hbar g_{A-cav} & 0 & \hbar\omega_A \end{pmatrix} \qquad (S3)$$

We diagonalize this matrix using QuTiP [s9] what let us obtain eigenstates and eigenvalues of this Hamiltonian. Values of energies of microcavity electromagnetic mode, donor and acceptor excitons were extracted from the experimental data. For finding the coupling strengths of our hybrid system we used differential evolution method [s10], stochastic method for obtaining extrema of function of several variables. In order to obtain correct experimental coupling strengths, we minimized the absolute value of difference between Hamiltonian eigenvalues and energies corresponding to experimental spectra maxima. As a result of this procedure we obtain the magnitudes of $g_{D-cav}$ and $g_{A-cav}$ that are equal to 435 meV and 41 meV.

In order to find the Hopfield coefficients, we obtained the eigenfunctions of Hamiltonian with coupling strengths corresponding to our experimental data. Every eigenfunction was represented as superposition of functions of pure photon and exciton states. The coefficients of the decomposition define Hopfield fractions of the polaritonic states.

**Supporting References**